\documentclass[]{spie}  
\usepackage[dvips]{graphicx}
\usepackage{textcomp,hyperref}
\usepackage{array,booktabs}
\usepackage{booktabs}
\usepackage{fancyhdr}
\usepackage{psfrag}
\usepackage[T1]{fontenc}
\usepackage{SIunits}
\usepackage{latexsym}
\usepackage{amsmath}
\usepackage{epsf}
\usepackage{float}
\usepackage{multicol}             
\usepackage{index}                
\usepackage{epsfig}
\usepackage{psfig}                
\usepackage{subfigure}
\usepackage[numbers]{natbib}

\def\aap{A\& A}

\def\araa{AnnRevA\& A}
\def\mnras{MNRAS}
\def\nat{Nature}
\def\apj{ApJ}

\def\apjl{ApJL}
\def\apjs{ApJS}

\makeatletter
\newcommand{\setfloattype}[1]{\renewcommand{\@captype}{#1}}
\makeatother

\title{X-RED: A Satellite Mission Concept To Detect 
\\Early Universe Gamma Ray Bursts} 

\author{Mirko Krumpe\supit{a}, Deirdre Coffey\supit{b}, \mbox{Georg Egger\supit{c},}  
		Francesc Vilardell\supit{d}, Karolien Lefever\supit{e}, Adriane Liermann\supit{f},
        Agnes I.D. Hoffmann\supit{g}, \mbox{J\"org Steiper\supit{h},} Marc Cherix\supit{i},
        Simon Albrecht\supit{j}, Pedro Russo\supit{k}, Thomas Strodl\supit{l},
 		Rurik Wahlin\supit{m}, Pieter Deroo\supit{e}, Arvind Parmar\supit{n},
        \mbox{Niels Lund\supit{o}, and G\"unther Hasinger\supit{p}}
\skiplinehalf
\supit{a}Astrophysikalisches Institut Potsdam, An der Sternwarte 16, D-14482 Potsdam, Germany; \\
\supit{b}The Dublin Institute for Advanced Studies, 5 Merrion Square, Dublin 2, Ireland; \\
\supit{c}Institute for Communication Networks and Satellite Communications, 
         \mbox{Graz University of Technology}, Inffeldgasse 12, A-8010 Graz, Austria; \\
\supit{d}Universitat de Barcelona, C/ Mart\'i i Franqu\`es 1, E-08028 Barcelona, Spain; \\
\supit{e}Instituut voor Sterrenkunde, Celestijnenlaan 200B, 3001 Leuven, Belgium; \\
\supit{f}Universit\"at Potsdam, Institut f\"ur Physik, Astrophysik, Am Neuen Palais 10, 
         \mbox{D-14469 Potsdam,} Germany; \\
\supit{g}Institut f\"ur Astronomie und Astrophysik, Universit\"at T\"ubingen, Sand 1, 
         \mbox{D-72076 T\"ubingen,} Germany; \\
\supit{h}Institut f\"ur Astrophysik, Universit\"at G\"ottingen, Friedrich-Hund-Platz 1, 
         \mbox{D-37077 G\"ottingen,} Germany; \\
\supit{i}Observatoire de Gen\`eve, Chemin des Maillettes 51, CH-1290 Sauverny, Suisse; \\
\supit{j}Leiden Observatory, P.O. Box 9513, NL-2300 RA Leiden, The Netherlands; \\
\supit{k}Faculty of Sciences - University of Porto, And Navegar Foundation, Av.24 n$^o$800, 
         \mbox{4500 - 202 Espinho,} Portugal; \\
\supit{l}Technische Universit\"at Wien, Karlsplatz 13, A-1040 Wien, Austria; \\
\supit{m}Department of Astronomy and Space Physics, Uppsala University, Box 515, 
         \mbox{SE-751 20 Uppsala,} Sweden; \\
\supit{n}Research and Scientific Support Dept. of the European Space Agency, ESTEC, Keplerlaan 1,
         2200 AG Noordwijk, The Netherlands; \\
\supit{o}Danish National Space Center, Juliane Maries Vej 30, 2100 Copenhagen, Denmark; \\
\supit{p}Max-Planck-Institut f\"ur extraterrestrische Physik, Giessenbachstrasse, Postfach 1312,
         \mbox{D-85741 Garching}, Germany; 
}
\authorinfo{Further author information: (Send correspondence to M. Krumpe)\\
M. Krumpe: E-mail: mkrumpe@aip.de, Telephone: +49 331 7499 328\\
\emph{published in}: Journal: Photonics for Space Environments X. Edited by Taylor, 
Edward W. {\em Proceedings of the SPIE}, Volume 58981J, (2005).
(c) 2005: SPIE--The International Society for Optical Engineering 
}

\begin{document} 
  \maketitle 

\begin{abstract}
Gamma ray bursts (GRBs) are the most energetic eruptions known in the
Universe. Instruments such as Compton-GRO/BATSE and the GRB
monitor on BeppoSAX have detected more than 2700 GRBs and, although 
observational confirmation is still required, it is now generally 
accepted that many of these bursts are associated with the collapse 
of rapidly spinning massive stars to form black holes.
Consequently, since first generation stars are expected to
be very massive, GRBs are likely to have occurred in significant numbers 
at early epochs. {\bf\em X-red} is a space mission concept designed to detect
these extremely high redshifted GRBs, in order to probe the nature of the
first generation of stars and hence the time of reionisation of the early
Universe. We demonstrate that the gamma and x-ray luminosities of typical
GRBs render them detectable up to extremely high redshifts ($z\sim10\,\mathrm{to}\,30$),
but that current missions such as HETE2 and SWIFT operate
outside the observational range for detection of high redshift GRB
afterglows. Therefore, to redress this, we present a complete mission
design from the science case to the mission architecture and payload,
the latter comprising three instruments, namely wide field x-ray
cameras to detect high redshift gamma-rays, an x-ray focussing telescope
to determine accurate coordinates and extract spectra, and an infrared
spectrograph to observe the high redshift optical afterglow. The mission
is expected to detect and identify for the first time GRBs with $z > 10$,
thereby providing constraints on properties of the first generation of
stars and the history of the early Universe.
\end{abstract}


\section{INTRODUCTION}
\label{sect:intro}

Before the launch of BeppoSAX in 1996, Gamma Ray Bursts (GRBs) were known only as explosive sources of gamma 
photons, without an identified counterpart in any other part of the spectrum. Subsequently, progress included 
the explanation of both temporal and spectral variations of the afterglow. However, although thousands of GRBs 
have now been discovered and observed, a large fraction remain unobserved at other wavelengths. The paucity of 
observations, poor positional accuracy and low photon fluxes have been important limiting factors. 

BATSE-observations of more than 1\,500 GRBs revealed two classes according to duration \cite{Balastegui2001}. 
GRB progenitor theory predicts that the class with the longest duration ($\sim$\,20\,to\,100\,s) is due to the 
collapse of massive stars \cite{Woosley1993}, \cite{Matheson2003}, and are detectable up to very large 
redshifts. For this class, Galama\,et\,al.\,(1998) \cite{Galama1998} provides strong evidence for a physical 
relation between GRBs and supernovae, which can be further classed as faint, low energy supernovae or bright, 
energetic ones \cite{Nomoto2005}. The latter, so called hypernovae, are associated with the death of fast 
rotating, massive stars \cite{Podsiadlowski2004}. Consequently, since early generations of stars are expected 
to
be very massive and fast rotators \cite{Abel2002}, \cite{Norman2000}, \cite{Marigo2003b},[10], 
\cite{Kudritzki2003}, GRBs are likely to have occurred in significant numbers at early epochs. Therefore, they 
are being increasingly recognized as potentially powerful probes of the very high redshift Universe 
\cite{Loeb2003}. 

Although several space missions have been launched to study GRBs (e.g. HETE2, INTEGRAL and SWIFT), extremely 
high redshift GRBs are likely to remain undetected/unidentified because their afterglow signatures are outside 
the observational range of current space missions. Only GRBs up to redshifts of z$\sim$6 can be observed by 
their optical instruments. However, luminosities of typical bursts render them detectable even as far as 
z\,$\sim$\,10\,to\,30 \cite{Lamb2000}. For this reason, we propose a space mission, {\bf\em X-red}, dedicated 
specifically to finding early Universe GRBs. Such bursts (z\,$\sim$10\,to\,30) will be redshifted into the 
x-ray region, with optical afterglows redshifted into the infrared region. This work is based on a preliminary 
mission study conducted by these authors at the ESA summer school on the ``Birth, Life, and Death of Stars'' 
at Alpbach, Austria, in 2004. {\bf\em X-red} is designed as an ESA project, to be launch between 
2012~and~2020, and is planned as a 3 year mission, with the possibility of extension up to 10 years. 

\section{Observable Science Data}
\label{sect:flux}

In the following, we examine the feasibility of detecting high redshift GRBs in terms of both statistical 
number of detections and also adequate flux levels at x-ray and infrared wavelengths. The results dictate the 
demands on instrumentation described in Section\,\ref{instruments}. Observations described in the literature 
of the Universe at redshifts (z\,$\sim$0\,to\,5) allowed us to use three important assumptions: GRBs are 
produced by hypernovae, GRBs have the same properties in the early Universe as they do today; and finally, 
flux extrapolation is possible to high redshifts. However, we are aware that there are several theories about 
the early Universe (z\,$\sim$5\,to\,30) that predict even higher GRB frequencies and fluxes. Therefore, it 
should be noted that our assumptions constitute a worst case scenario.

We can gauge the expected number of detectable GRBs from the detection rate of SWIFT \cite{Gorosabel2004} 
combined with our chosen instrument field of view (FOV) and sensitivity, Section\,\ref{WFC_fig}. The detection 
rate of {\bf\em X-red} is therefore estimated from Fig. 3 of Hartmann et al. (2004) \cite{Hartmann2004}, 
given the higher sensitivity of {\bf\em X-red} compared to SWIFT. The total number is
expected to be on the order of 800 detections over the 3 year mission lifetime. Of these detections, around 
10\,\% are expected to be from GRBs at z\,$>\,$5 (i.e. at least 80 GRBs at a distance beyond which GRBs have 
not yet been observed) while 2\,\% (i.e. $\sim$15 detections) are expected to be from GRBs at z\,$>\,$10. 

\begin{figure}[ht]
\centering
\includegraphics[width=7.5cm]{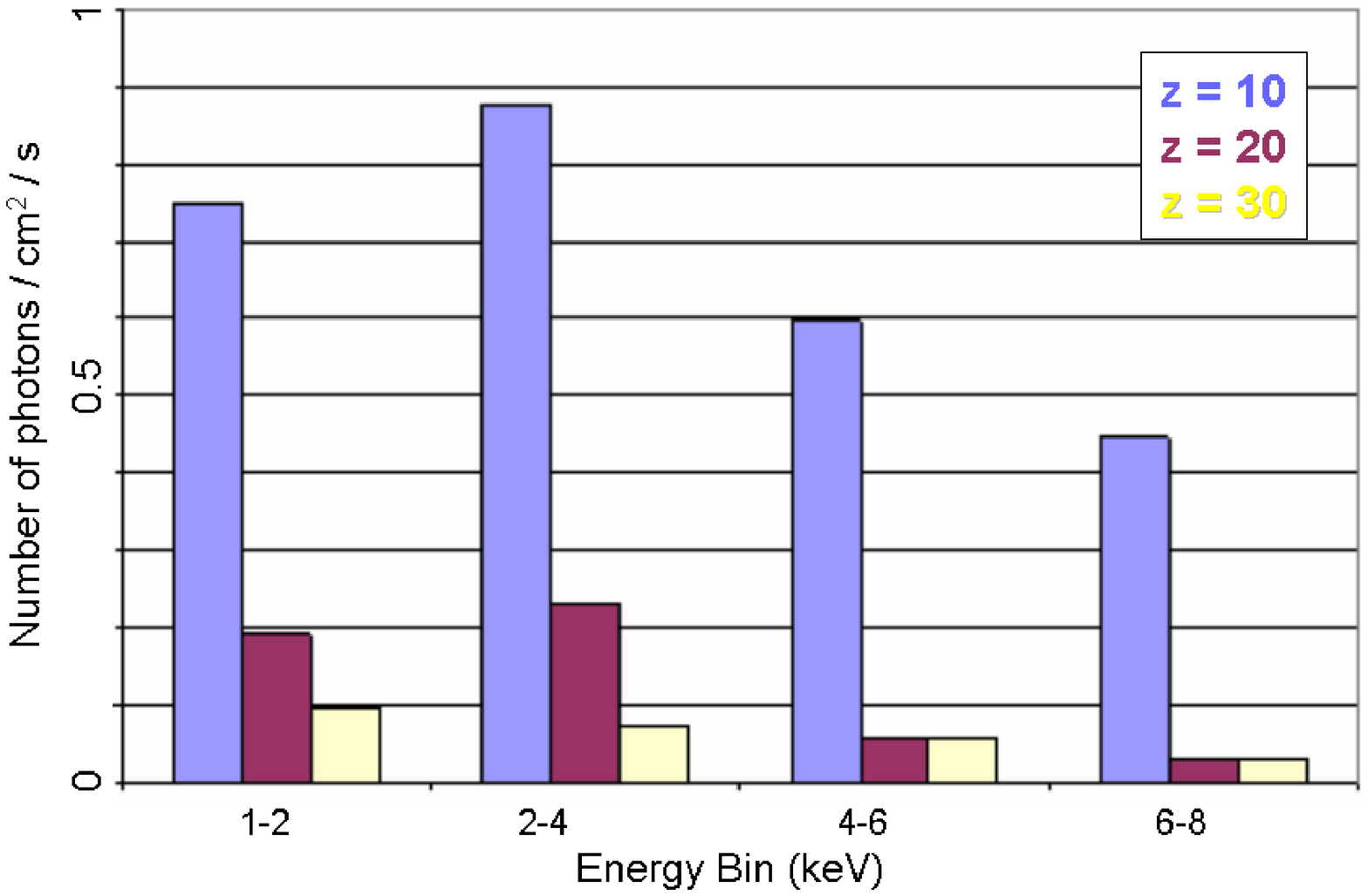}
\hspace{1.0cm}
\includegraphics[width=5.5cm]{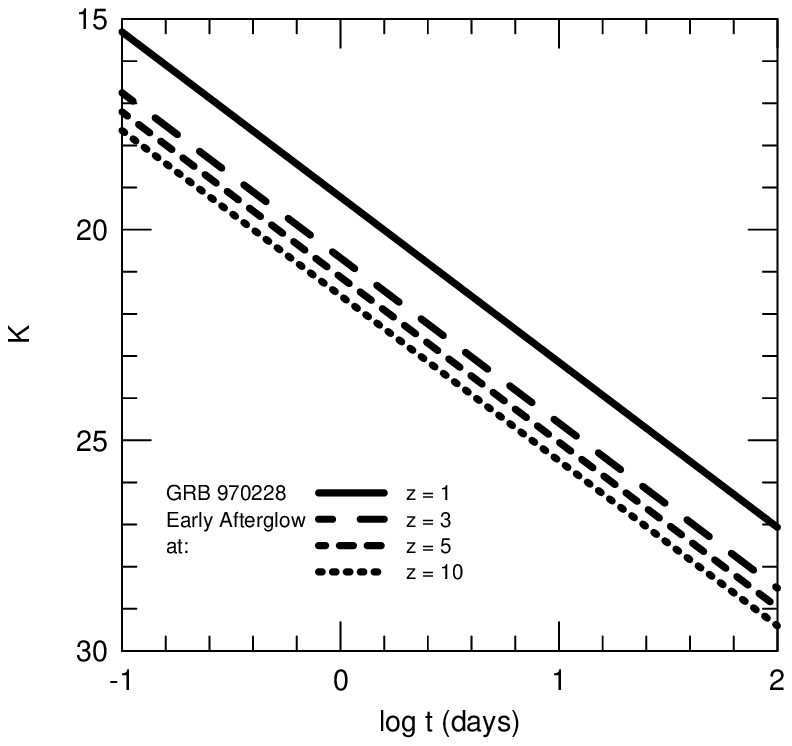}
\caption{
Expected peak emission and afterglow fluxes of GRBs. 
{\bf\em Left panel:} Prompt emission photon flux from GRBs at various redshifts. 
{\bf\em Right panel:} Best-fit spectral flux distribution of the afterglow of GRB970228 from \cite{Lamb2000}. 
\label{fluxboys}
}
\end{figure}

For high redshifted GRBs (z\,$>$\,10) the peak energy output of 100\,keV to 600\,keV is shifted into the x-ray 
regime. The observed luminosities span a few orders of magnitude, but tend to higher values at higher 
redshifts 
\cite{Yonetoku2004}. At z\,$\sim$\,10, the expected luminosities are a few times 
10$^{53}$-10$^{54}$\,erg\,cm$^2$\,s$^{-1}$. The expected photon number at a given energy in the GRB spectrum 
by Meegan (1998) \cite{Meegan1998} was integrated over the appropriate energy range. Additionally, due to the 
time dilatation, the GRB is extended to z+1, and so a burst that lasts for $\sim$\,10\,s will be observable 
for $\sim$\,100\,s. This provides the time constraint within which the spacecraft must point its follow-up 
instruments. Unfortunately, however, the time dilatation also causes a decrease in the flux rate of high 
redshifted GRBs. Figure \ref{fluxboys}, left panel, takes that into account and shows the calculated flux 
distribution assuming a luminosity of 10$^{54}$\,erg\,cm$^2$\,s$^{-1}$. 

Although the number of photons increases at lower energy bins, so too does the diffuse x-ray background. 
Therefore, the FOV must be limited in order to detect the GRB above the background. The x-ray afterglow is one 
to two orders of magnitude lower than the actual burst \cite{vanParadijs2000}. Therefore, a {\em focussing} 
x-ray telescope is needed to study x-ray afterglows, or even late x-ray bursts, taking place several minutes 
after the prompt emission \cite{Piro2005}. X-ray focusing telescopes work in an energy range of 0.1-12\,keV. 
Therefore, a possible absorption due to photo-ionization, which is highest at low energies and drops with the 
third power of energy, is now investigated. In {\bf\em X-red}'s observation field, near the galactic poles, 
the hydrogen number density is N$_{\mathrm{H}}$~=~10$^{20}$cm$^{-2}$ \cite{Stark1992}. According to 
Wilms\,et\,al.\,(2000) \cite{Wilms2000} the absorption at 0.1 keV is 4\,\% and decreases fast with energy. 
Thus, absorption is negligible. 

In the infrared wavelength range, the dimming of the afterglow, redshifting, and time dilatation nearly cancel 
one another out \cite{Lamb2000}. The extinction at high redshifts is likely to be small because of the rapid 
decrease in metallicity beyond z$\sim$3. Figure\,\ref{fluxboys}, right panel, shows the time dependant 
decrease of afterglow flux. It can be extrapolated to a short time after the burst (Table\,\ref{flux_K}) and 
specifies the demands on the infrared telescope.

\begin{table}[H]
\scriptsize{
\begin{center}
 \begin{tabular}{cc} \toprule
  Time after burst 	&K-band Flux 	\\ 
	(s)		&(Jy)		\\ \midrule
  10 & 2.5 \\
  60 & 0.15 \\
  180 & 0.026 \\
  360 & 0.0086 \\ \bottomrule
 \end{tabular}
\caption{Extrapolated afterglow fluxes in the K band}
\label{flux_K}
\end{center}
}
\end{table}
\newpage
\section{Instrumentation} 
\label{instruments}

\begin{figure}[ht]
\centering
\includegraphics[width=6cm]{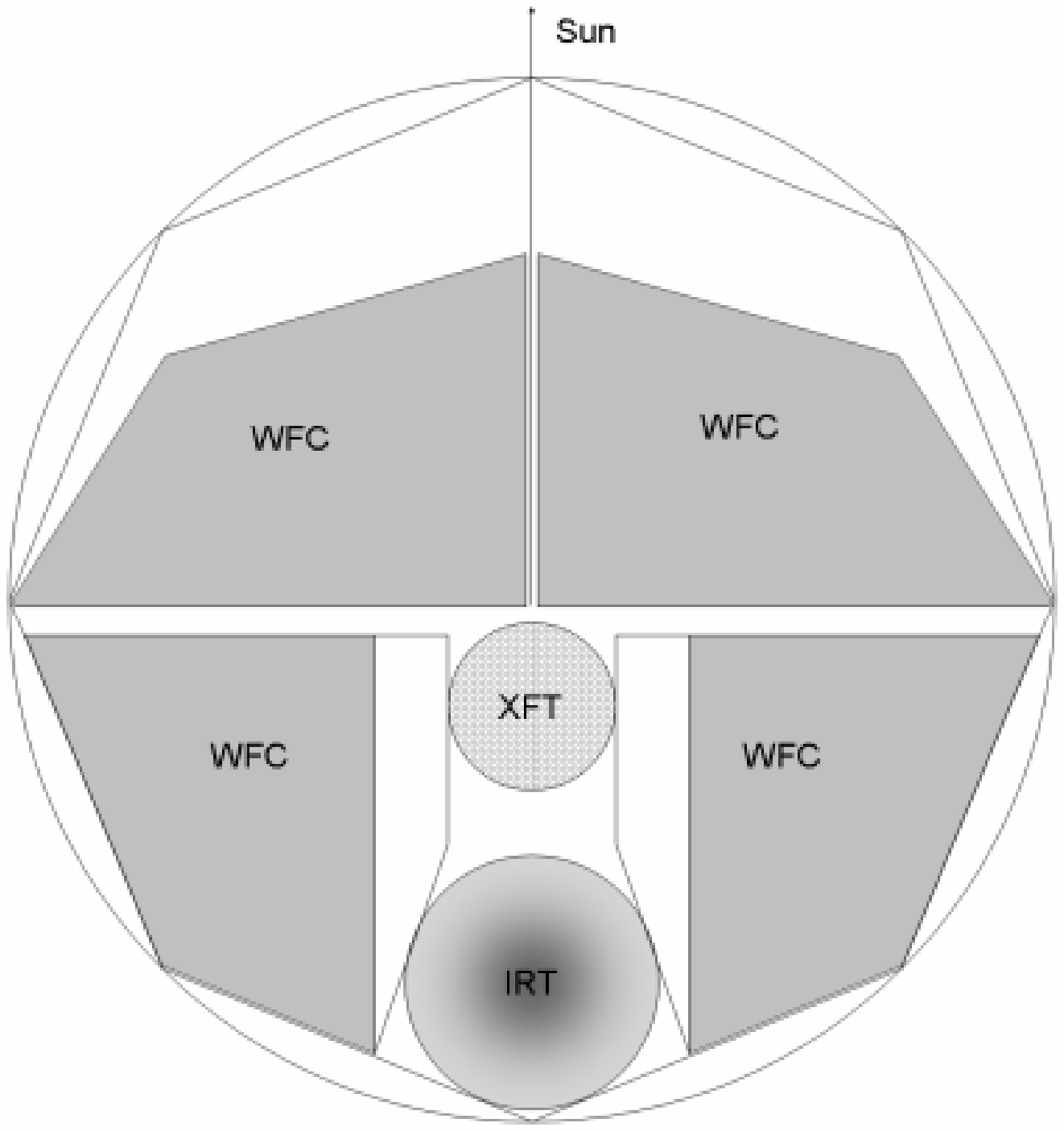}
\hspace{1.5cm}
\includegraphics[width=5cm]{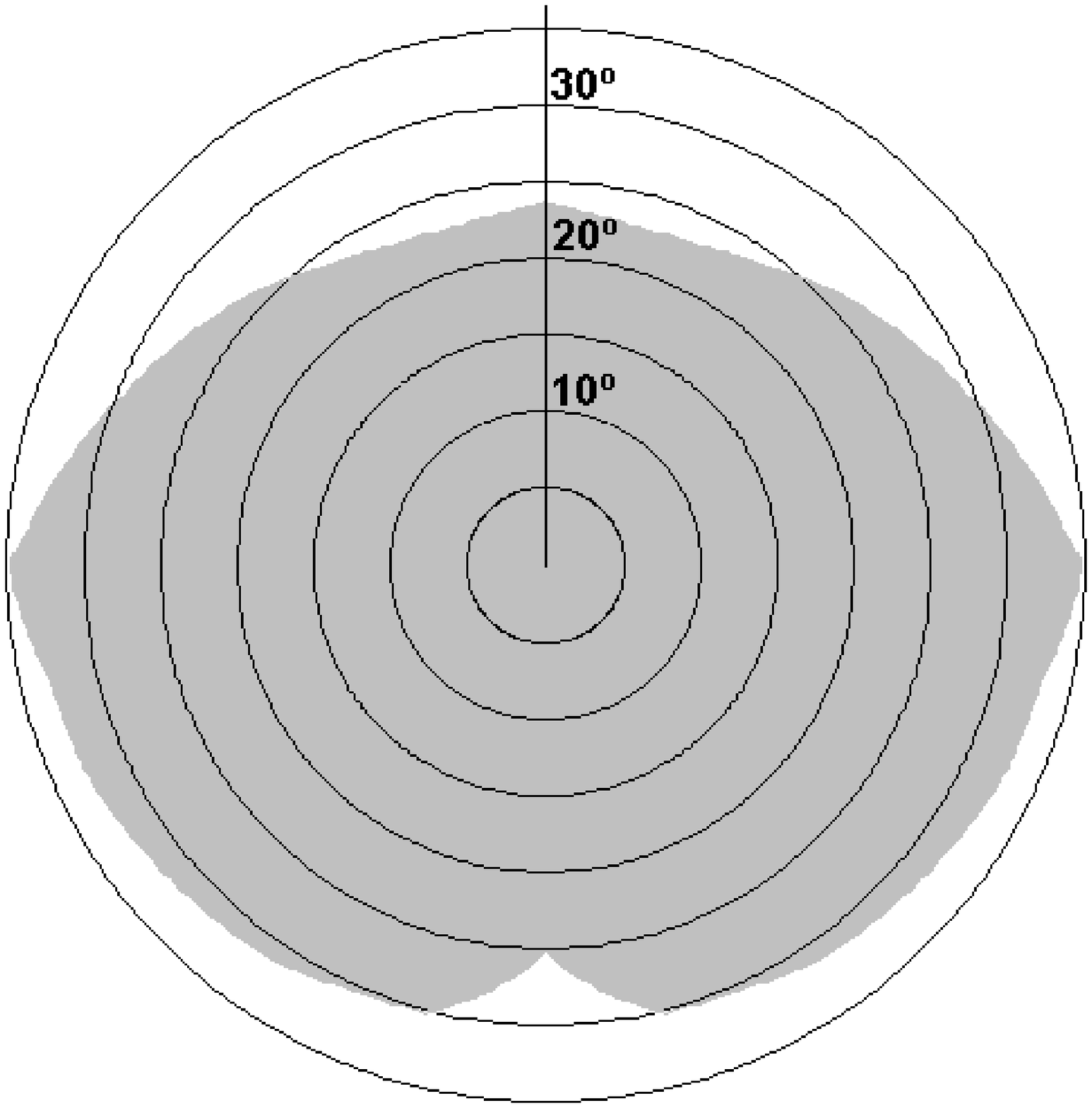}
\caption{{\bf\em Left panel:} Layout of the three different instrument types on board the {\bf\em X-red} 
satellite. 
{\bf\em Right panel:} Field of view of the four WFCs. The radial axis is pointing towards the Sun. 
\label{xred_instruments}}
\end{figure}

The main objective of {\bf\em X-red} is to detect a reasonable number of high-redshifted GRBs with a high 
positional accuracy. To achieve this objective, the satellite will be equipt with three types of instrument 
(see layout in Figure\,\ref{xred_instruments}, left panel). Four Wide Field Cameras (WFCs) operating at x-ray 
wavelengths will detect high redshift gamma-rays and obtain rough coordinates. The X-ray Focussing Telescope 
(XFT) will perform imaging observations to improve coordinate accuracy, and will also extract x-ray spectra 
from which redshifts may be determined. The InfraRed Telescope (IRT) will extract near infrared spectra of the 
high redshift afterglow. 

\subsection{Wide Field X-ray Cameras}
\label{WFC_fig}

WFCs operating at x-ray wavelengths will detect high redshift gamma-rays and obtain rough coordinates. The 
only instruments currently available for observing a large FOV in the x-ray region are coded mask detectors. 
Each WFC consists of a coded mask placed 1.67\,m above the detector plane
by a rigid composite structure. The coded masks, e.g. \cite{Vigneau2003} of $3\times3\times1$\,mm
(L$\times$W$\times$H) lead pieces, were
designed to cover the whole available surface of the service module. 
The
dimensions of the satellite structure and the sun~shield put engineering
constraints on the design and placement of the WFCs. For this reason, there are
two pairs of masks (see Figure\,\ref{xred_instruments}, left panel). The two masks closer to the Sun are 
symmetric, have a maximum size of $1.71\times1.18$\,m and have almost 160\,000 elements each (with $\sim50$\% 
of them transparent to
x-rays). The two masks on the further side from the Sun, also symmetric, have a
maximum size of $1.41\times1.40$\,m and over 161\,000 elements each. 

The two main problems posed by detectors covered by coded masks are the background photon noise level and the 
lack of collimation in photons reaching the detector itself. Since {\bf\em X-red} requires not only a large 
FOV but high sensitivity (see Section\,\ref{sect:flux}), these difficulties must be overcome. To overcome the 
background photon noise, a combination of four WFCs is used, with each WFC surrounded by a shield. More 
importantly, it is now possible to avail of the Depleted Field Effect Transistor (DEPFET) detector technology, 
a new CCD technology with hybrid soft and hard x-ray detection and high speed individual pixel readout, 
currently in development for the XEUS mission. This technology allows considerable improvement in background 
noise reduction, see below. (The lack of flux collimation is addressed by the XFT design, Section\,\ref{xrt}.)

Each WFC is equipped with two 70$\times$70\,cm detectors, covering an energy range between 0.1\,keV and 
100\,keV. A 70$\times$70 cm Depleted Field Effect Transistor (DEPFET) is placed in front of a 70$\times$70\,cm 
cadmium telluride (CdTe) detector, such that hard and soft x-ray detection is possible (see \cite{Bavdaz2004a} 
for a complete explanation of DEPFET and CdTe detectors). The two detectors can be used to trace the photon's 
path. By knowing the detection position in both detectors, it is possible to eliminate background photons. 
Hence, the background photon noise can be reduced even further. 

\begin{table}[!hbt]
\scriptsize{
 \begin{center}
  \begin{tabular}{llr} \toprule
   & Soft x-ray detector & Hard x-ray detector \\ \midrule
   Size &  70$\times$70 cm$^2$ & 70$\times$70 cm$^2$ \\
   Number of elements & 700$\times$700 & 700$\times$700 \\
   Size of each element & 1$\times$1 mm$^2$ & 1$\times$1 mm$^2$ \\
   Spectral range & 0.1 - 10 keV & 10 - 100 keV \\
   Power consumption & 20 W & 25 W \\ \midrule
   Total WFCs Mass & \multicolumn{2}{c}{4 x 40 kg}\\
  \bottomrule
  \end{tabular}
  \caption{Estimated properties for the WFCs detectors.}
  \label{WFCdettable}
 \end{center}
}
\end{table}

High redshifted GRBs ($z\sim 20$) are fainter than medium redshifted GRBs 
($z\sim 10$). Coded mask instrument sensitivity increases with the size of 
the detectors. According to the expected fluxes, 
all detectors have been designed large enough to be able to detect a typical GRB
at $z=20$ within their energy range. The large detector dimensions and the number and size of the pixels (see 
Table\,\ref{WFCdettable}) may be considered ambitious. However, it is expected that technological advances are 
likely to meet this demand by the planned launch date of {\bf\em X-red}. 

Overall, with this configuration, the total FOV of the four WFCs (which have coded masks of 50\,\% 
transparency) is 0.6\,steradians ($\sim\,70\,^{\circ}\,\times\,50\,^{\circ}$, Figure\,\ref{xred_instruments}, 
right panel), and the positional accuracy is $\sim$\,1\,arcmin (FWHM\,$\sim$\,6\,arcmin), as determined by the 
mask-to-detector distance, the size of the mask elements and the detector pixels.

\subsection{X-Ray Focussing Telescope}
\label{xrt}

The second x-ray instrument on-board is the XFT, which will perform imaging observations to improve coordinate 
accuracy, and will also extract x-ray spectra from which redshifts may be determined. 
With this instrument, we can overcome the second difficulty posed by the WFC coded mask (mentioned in 
Section\,\ref{WFC_fig}), i.e. the lack of collimation in photons reaching the detector. 

X-ray grazing incidence mirrors are able to focus x-rays but only under very small reflection angles and, with 
increasing photon energy, the angle becomes even smaller. This results in the typically large focal lengths of 
x-ray telescopes. Meanwhile, the dimensions of the fairing (Section\,\ref{engineering}) and moment of 
inertia considerations (Section\,\ref{attitude}) impose a limit on the available focal length for the XFT of 
5.5\,m, with corresponding radius 0.28\,m (scaling a XMM-Newton type telescope). We must look to new 
technologies to ensure adequate sensitivity under these focal length and radius restrictions.

Silicon High-Precision Pore Optics (Si-HPO, or simply `pore optics') is another challenging new technology 
currently in development for the XEUS mission (as is DEPFET, Section\,\ref{WFC_fig}). Although the constraint 
on maximum tolerable graze angle for pore optics is the same as for conventional mirrors, pore optics can be 
made to more efficiently fill the aperture of the telescope. This means it is possible to obtain a given 
effective mirror area with a telescope of smaller opening. This in turn translates to a shorter focal length. 
So the use of pore optics will not only have the effect of drastically decreasing the mass of the mirrors 
since it is very light-weight, but will increase the effective area substantially. With our chosen design, a 
pore optics telescope will have an effective area of 1\,400\,cm$^2$\,@\,1.5\,keV, which is 12 times better 
than SWIFT \cite{Bavdaz2004}, \cite{Beijersbergen2004}. 

Having collimated the photons, we then choose to implement a photon counting intrinsic energy resolution 
detector. As with the WFCs, we use DEPFET detector technology of $3.2~\mathrm{cm}\times 3.2~\mathrm{cm}$ with 
$640 \times 640$ pixel. The detector will be read out every 1\,ms in window mode (i.e. only partial detector 
read-out according to GRB location, resulting in increased speed and hence allowing photon energy resolution). 
The observational data will yield {\em both} x-ray images and spectra. Given a FOV of 10\,arcmin a GRB 
position accuracy of 1\,arcsec is feasible. Further technical details are listed in Table\,\ref{tab:XPTdata}. 
From the x-ray spectrum, we can identify emission and/or absorption lines which allow determination of the 
redshift. 

\begin{table}[!hbt]
\scriptsize{
  \begin{center}
 \begin{tabular}{llr} \toprule
  &Telescope & Detector \\ \midrule
  Type & Si-HPO & DEPFET \\ 
  Dimensions & r = 28 cm & 3.2 cm $\times$ 3.2 cm\\
 & F = 5.5 m & 640 $\times$ 640 pixel with 50$~\mu$m $\times$ 50$~\mu$m \\ 
 Mass & 50 kg & 40 kg \\ \midrule
 Resolution & \multicolumn{2}{c}{mirror 5" $\Rightarrow$ GRB position: 1"}\\ 
 Energy range & \multicolumn{2}{c}{$0.1-10$ keV} \\
 & \multicolumn{2}{c}{ 150 eV FWHM}\\ 
 Effective Area & \multicolumn{2}{c}{$A_{\mathrm{eff}} = 1400 ~{\mathrm{cm}^{2}}$ @ $1.5~{\mathrm{keV}}$}\\ 
 Field of View & \multicolumn{2}{c}{10 arcmin} \\ 
 \bottomrule
 \end{tabular}
  \end{center}
  \caption[tab:XPTdata]{\label{tab:XPTdata} Technical details for the XFT.}
}
\end{table}

\subsection{Infrared Telescope}

The IRT is intended to extract spectra of the high redshift optical afterglow. The demands on the IRT are 
dictated by the nature of high redshifted afterglows of GRBs. These events are limited in lifetime (days to 
months) and the flux is continuously decreasing rapidly, so it is crutial to obtain a spectrum as soon as 
possible.

In the majority of GRB afterglows, Lyman-$\alpha$ is the most distinctive absorption line \cite{Vreeswijk2004} 
and is detectable even with low-resolution spectroscopy and in a noisy spectra. Even though other emission 
and/or absorption lines are recognized in afterglows, we expect Lyman-$\alpha$ to be the most prominent line 
from the first generation of stars, due to their chemical composition. At a redshift of $\sim$5\,to\,15 or 
even up to z$\sim$20, Lyman-$\alpha$ is shifted into the near infrared 
($\lambda$\,$\sim$\,0.73\,to\,1.95\,$\mu$m, or even \mbox{2.55\,$\mu$m}). However, for clear determination of 
the redshift, detection of other features is also desirable. Hence, the detector's wavelength range should 
exceed $\lambda\,=\,4.0\,\mu$m. For an adequate signal-to-noise ratio, our choice of detector must operate at 
$\sim$35\,K. The corresponding demand on thermal design puts strong constraints on the telescope, the detector 
and the whole spacecraft. Active cooling systems have the disadvantage that they limit the lifetime of the 
mission significantly and increase the spacecraft's mass substantially. Passive cooling to 35\,K is an 
engineering challenge but it is absolutely desired, and possible if the complete satellite architecture is 
designed according to this requirement (Section\,\ref{thermal_design}). 

Considering the required spectral range and the feasibility of passive cooling, the Near Infrared Spectrograph 
(NIRSpec) of the James Webb Space Telescope can serve as a basis for the detector design 
\cite{Hofferbert2004}. This operates in the wavelength range of 0.6\,to\,5.0\,$\mu$m and offers a range of 
resolutions (R\,=\,100\,-\,3\,000). Given that afterglow fluxes decrease rapidly over time, high resolution 
spectroscopy is not possible since the photon rates are low and space-based mirrors are too small. However, 
for determination of redshift, low resolution spectroscopy is sufficient. Therefore, we propose a modified 
NIRSpec with a "low" resolution mode of R\,$\sim$\,30 and a "high" resolution mode of R\,$\sim$\,100. For a 
telescope diameter of 0.85\,m the required minimum near infrared fluxes for a 10\,$\sigma$ event corresponds 
to 8.6\,(or 28.5)\,mJy (by scaling NIRSpec) at a given resolution of R\,$\sim$\,30 (or R\,$\sim$\,100).

\begin{table}[!hbt]
\scriptsize{
 \begin{center}
  \label{infrared}
  \begin{tabular}{llr} \toprule
    & IR-Telescope & IR-Detector \\ \midrule
    Diameter &  0.85 m & 0.80 m   \\
    Height   &  1.5 m  &          \\
    Operating temperature& 50 K & 35 K\\
    Mass & 50 kg & 100 kg\\   \midrule
    Spectral range & \multicolumn{2}{c}{0.6 - 5 $\mu$m} \\
    Resolution &  \multicolumn{2}{c} { R $\sim$ 30, R $\sim$ 100}\\
    Field of View &  \multicolumn{2}{c} { 3.4 $\times$ 3.4 arcmin}\\
    Power consumption & \multicolumn{2}{c}{ 20 W }\\ \bottomrule
   \end{tabular}
\caption{Estimated properties for the IRT and the infrared detector.}
 \end{center}
}
\end{table}

\section{Mission Architecture}
\label{architecture}

\subsection{Orbit \& Launcher}
\label{orbit}

The spacecraft orbit was chosen as the Lagrangian point, L2, where gravitational pull and centripetal force 
between Sun and Earth are equal. This location will provide long term uninterrupted observations of GRBs and 
their afterglows with the XFT and IRT, which would not be achieveable from an Earth orbit. The location will 
also avoid eclipses, giving a stable thermal environment. 

Thanks to the advanced light-weight instrument technology (Section\,\ref{instruments}) and choice of 
light-weight passive cooling (Section\,\ref{engineering}), the mass loading is minimised and so we can choose 
the cheaper Soyuz launcher, which imposes a lower mass constraint than the more expensive Ariane 5. This choice also, 
however, gives some limiting constraints to the total dimensions of the spacecraft. It will be equipt with the 
new ST fairing (offering larger dimensions for a bigger payload). The internal dimensions of this fairing are 
$\sim$3.8\,m in diameter and $\sim$9.5\,m in height, but forming a cone at the top which 
leaves an effective height H$_{eff}$$\sim$\,6\,m for the basic 3.6\,m~diameter cylindrical shape. A Fregat module, which 
provides the final injection to L2, is included in the fairing. 

It is expected that the Soyuz--Fregat spacecraft will be launched from the European Spaceport in Korou. The 
Soyuz will bring the spacecraft to an intermediate orbit around Earth at which point the Fregat module then 
performs the injection to the transfer orbit, including a fly-by maneuver with the moon, until it finally 
reaches L2. Upon arrival, the spacecraft will perform a halo--orbit about the instable L2 point, with a total 
amplitude of $8\times 10^8$\,m. 

\subsection{Propulsion}
\label{propulsion}

In order to correct the flight trajectory to L2, and to maintain stability about the instable L2 point, a 
propulsion system will be included in the spacecraft. Additionally, this will serve to offload the momentum 
wheels (Section\,\ref{attitude}), which are included in the spacecraft for pointing manoeuvres. Required 
trajectory corrections of 50\,m\,s$^{-1}$ and stability corrections of 2\,m\,s$^{-1}$\,yr$^{-1}$ are expected. 
Estimating an extended maximum lifetime of 10\,years for the mission, the sum of all corrections is 
$\sim$\,70\,m\,s$^{-1}$. 

The thruster propellant chosen is hydrazine (N$_2$H$_4$), for its high specific impulse (I$_{sp}$\,$\sim$\,230\,s) 
compared to a liquid gas like nitrogen (I$_{sp}$\,$\sim$\,70\,s). Electrical thrusters (I$_{sp}$\,$>$\,1\,000\,s) do 
exist but they require a stable high voltage supply and a non-standard power processing unit, the mass of 
which increases with specific impulse. Also, they offer best performance for long thrust periods, and so are not 
required for this mission. Furthermore, structure and mass constraints preclude electrical thrusters. 
Therefore, hydrazine thrusters are favoured, as their high specfic impulse means lower fuel mass rendering them more 
suitable for longer duration missions. In addition to the propellant mass, the specifications for the 
thrusters, pipes and valves have to be considered. For the expected requirements, a system of 16 thrusters 
each providing a force of 10\,N is deemed sufficient. The propellant will be stored in two tanks situated in the service 
module. The mass estimated for a total lifetime of 10\,years is $\sim$\,70\,kg, and is broken down as follows: 
propellant $\sim$50\,kg; tanks (2$\times$6\,kg) $\sim$12\,kg; thrusters (16$\times$0.3\,kg) $\sim$4.8\,kg; 
pipes and valves $\sim$3\,kg. 
  
\subsection{Attitude Control} 
\label{attitude}

Since using thrusters for the attitude control would add extra propellant mass and exceed the mass limitations 
of the spacecraft, we choose instead a four momentum wheel system (for redundancy purposes) orientied in
a tetrahedron to control the rotation of the spacecraft. This attitude control system will 
include two star trackers, with a positional accuracy of 0.25\,arcsec (FWHM$\sim$1.0\,arcsec), pointing parallel to the 
XFT -- IRT alignment. The spacecraft is of effective height H$_{eff}$\,$\sim$\,6\,m (Section\,\ref{orbit}) and of 
mass m\,$\sim$\,1\,500\,kg (Section\,\ref{mass_estimate}). Assuming a cylindrical geometry, the satellite's 
highest possible rotational moment of inertia, $I$, (i.e. about the x or y axis, see 
Figure\,\ref{fig:sc_geometry}) was estimated by 
\begin{eqnarray}
I = \frac{1}{12}\,m\,(H_{eff}^2+3\,R^2) + m\,H_{com}^2 \nonumber
\label{MOI}
\end{eqnarray}
where R is the spacecraft radius and H$_{com}$ is the height of the center of mass of the cylinder 
(H$_{com}$=$\frac{1}{2}$\,H$_{eff}$-H$_{SM}$, see Section\,\ref{engineering}). The first term represents a 
rotation around the center of the cylinder. The second term is a correction term, using Steiner's law, to 
account for the distance offset between the center of the cylinder and its center of mass, since the center of 
mass is located between the service module (SM) and payload module (PM) (Figure\,\ref{fig:sc_geometry}, left 
panel). The resulting moment of inertia is $I$\,$\sim$\,9\,100\,kg\,m$^{2}$. Since the momentum wheels must slew 
the satellite to any point in the FOV of the WFCs within 60 seconds, an angular speed of 
$\sim$0.01\,rad\,s$^{-1}$ must be achieved. Therefore, the momentum wheels are required to store an angular 
momentum of 91\,N\,m\,s. 

\begin{table}[ht]
\scriptsize{
\centering
\begin{tabular}{lr}\toprule
Angular momentum & 91 Nms\\
Operating speed &       8000 rpm\\
Max. speed      &      15000 rpm\\
Weight          &       4 x 11 kg = 44 kg\\
Size(diameter, height) &    280 mm x 30 mm\\
Peak power      &       750 W\\
Mean power      &       100 W\\ \bottomrule
\end{tabular}
\label{momentumwheelspecifications}
\caption{Specifications of the momentum wheels (based on 
http://www.vfct.com/satellites/wheels)}
}
\end{table}

\subsection{Spacecraft Engineering}
\label{engineering}

\begin{figure}[ht]
\centering
\includegraphics[width=4.7cm]{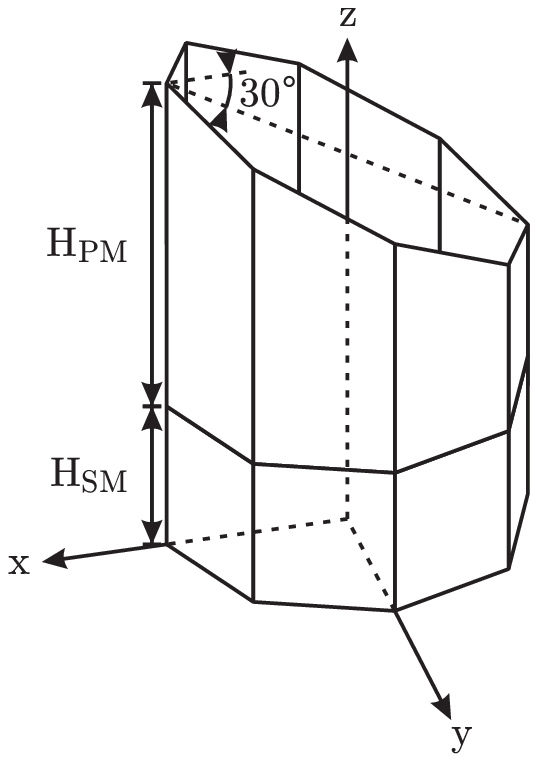}
\hspace{1.9cm}
\includegraphics[width=4cm]{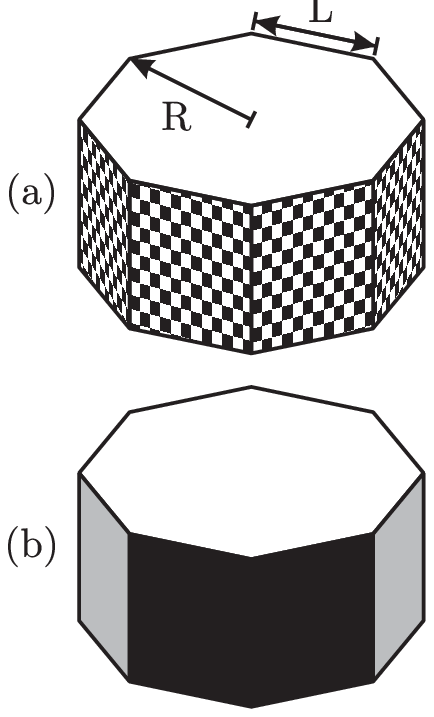}
\caption{Geometry of the spacecraft. {\bf\em Left Panel:} The geometry of the division into SM and PM 
($H_\mathrm{PM}$=4.5~m, $H_\mathrm{SM}$=1.5~m). {\bf\em Right Panel:} The geometry of the sun-facing front (a) 
and dark rear (b) of the SM ($R$=1.8~m, $L$=1.34~m).
\label{fig:sc_geometry}}
\end{figure}

The spacecraft is divided into two parts: the service module (SM) and the payload module (PM) 
(Figure\,\ref{fig:sc_geometry}, left panel). The PM carries all the scientific instruments, the communication
antenna, star trackers, sun shields and passive cooling radiators. The SM includes all the 
devices necessary for housekeeping, attitude control, propulsion system, etc. The solar panels are mounted on 
the side facing the Sun (chequered areas in Figure \ref{fig:sc_geometry}(a). The back side is equipped with 
black painted radiatiors (Figure \ref{fig:sc_geometry}(b)). Finally, due to its comparatively long focal length, the XFT 
necessarily extends into the SM. 

General considerations in spacecraft engineering include the structure and thermal design. The structure must provide the 
stiffness necessary to survive the launch, and to assure pointing accuracy during the mission. The thermal design must 
ensure the instruments
are shielded from solar radiation, since they need to be kept at certain operational temperatures. 
Rather than choosing a cylindrical shape, an octagonal structure for the spacecraft was chosen for ease of design 
(particularily regarding the solar panels) and cost efficiency. Lastly, the surface of the spacecraft has to 
be electrically conductive, avoiding potential differences that could cause discharges and damage to the 
devices on-board. 

\subsubsection{\label{sect:powersystem}Power Systems}

\begin{figure}[ht]
        \begin{minipage}[b]{.4\textwidth}
			\setfloattype{table}
           \begin{tabular}{lr} \toprule
                Subsystem & Power requirement \\ \midrule
                WFC & 180~W\\
                IRT & 20~W\\
                XFT & 15~W\\
                Momentum wheels & 100~W, (750~W peak)\\
                Telecommunication & 100~W\\
                Data handling & 20~W\\
                Attitude sensors & 20~W\\
                Power system itself & additional 20\%\\ \midrule
                Total & ca. 550~W \\ \bottomrule
             \end{tabular}           
          \caption{Estimated power requirements\label{tab:powerrequirements}}
        \end{minipage}
\hfil
        \begin{minipage}[t]{.4\textwidth}
          \includegraphics[width=7cm]{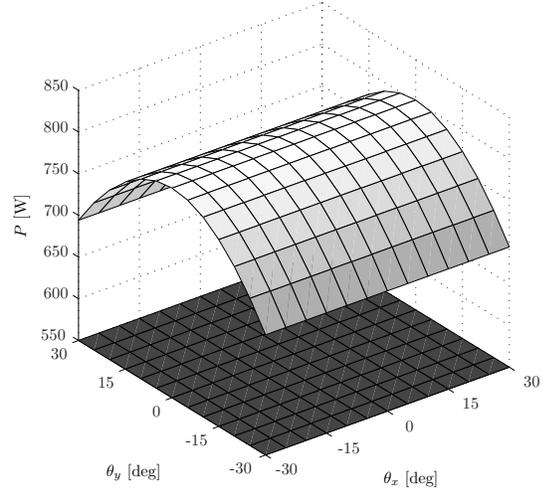}
          \caption{Required and obtained electrical power from the solar panels.}
          \label{fig:psolar_xy_final}
        \end{minipage}
\end{figure}

The major components of the power systems consist of solar arrays, storage batteries, a power regulator system 
and a power distribution system. In this section we will focus on the main power source, the solar arrays. The 
orbit for {\bf\em X-red} around L2 will avoid eclipses, thus the solar arrays are a reliable source of power. 
However, storage batteries must be included as well, in order to provide peak power to the spacecraft and 
also as a backup power source for possible emergency scenarios. To design the solar arrays, the power 
requirements of the different subsystems were estimated (Table\,\ref{tab:powerrequirements}). 

The power loss from the solar arrays to the loads is expressed by the supply efficiency $\eta_s$\,$\sim$\,0.85, 
while further degradations (design and assembly inefficiencies, degradations due to non-optimal temperature of 
the array and inherent degradation) add up to a deployment efficiency $\eta_d$\,$\sim$\,0.55 \cite{SMAD}. To 
fulfil the power requirements, highly efficient solar arrays are mounted on the SM. A product from Spectrolab 
was chosen, for illustration. The Ultra Triple Junction~(UTJ) Solar Cells have a minimum average efficiency at 
the beginning of life~(BOL) of $\eta_{BOL}$\,=\,28\,\% and at the end of life~(EOL) 
$\eta_{EOL}$\,=\,24.3\,\%. The solar radiation density (solar constant) at the L2~orbit (i.e. at a distance 
of 1\,AU\,+\,1.5\,$\times$\,10${^9}$\,m from the sun) is $J_{_\mathrm{L2}}\,=\,1344~\frac{W}{m^2}$. The power output from 
a solar cell at the EOL (worst case) is then controlled by the following relationship: 
$$P_{\mathrm{solar}} = J_{_\mathrm{L2}} ~ \eta_{\mathrm{EOL}} ~ A ~ \eta_s ~ \eta_d \cos\theta$$ 
where $\theta$ is the solar zenith angle (i.e. the angle between the normal vector of the area and the solar rays) and $A$ 
is the area of the cells. By simple considerations, the required area amounts to 3.5\,m$^2$. However, since our spacecraft 
is octagonal and our
spacecraft is designed to rotate by $\pm\,30^{\circ}$ about the x and/or y axis (implying that sides facing the sun always have different solar zenith angles), a more sophisticated approach is desirable. A 
simulation program was written to study in detail the impact of rotation on power generation and thermal 
aspects. As a result of the geometry of the SM, the total area of the solar panels is $4~\mathrm{m}^2$. 
Figure \ref{fig:psolar_xy_final} illustrates that the chosen area guarantees the required electric 
power even given EOL and low light conditions. 

\subsubsection{Thermal design}
\label{thermal_design}

When faced with a long mission duration and mass limitations, passive cooling becomes imperative. The primary 
engineering challenge is the cooling of the IRT, which will operate at 35-50\,K. The spacecraft will orbit around L2, thus one 
side will always point towards the Sun and the other always away from it. A black
body in space will radiate according to the Stefan Boltzmann fourth power law, and will absorb the incident solar 
radiation $J_{\mathrm{incident}}$: 
$$ J_{\mathrm{radiated}}=\varepsilon \sigma T^4\ ,\ \ \ \ J_{\mathrm{absorbed}}=\alpha J_{\mathrm{incident}}$$
where $\varepsilon$ is the emittance, $\alpha$ is the absorptance, and $\sigma$ is the Stefan-Boltzmann 
constant equal to $5.67\times10^{-8} \frac{~\mathrm{W}}{\mathrm{m^2K^4}}$ \cite{SSE}. Choosing materials with 
the right absorptance and emittance properties is crucial. The sum of the absorbed power and dissipated power should not 
exceed the radiated power. A spacecraft in thermal equilibrium fulfils 
$$\underbrace{J_{_\mathrm{L2}} \sum_{i} \alpha_i A_i \max\{0, \cos
  \theta_i\}}_{P_\mathrm{absorbed}} + P_{\mathrm{dissipated}} =
\underbrace{\sigma \sum_{j}  \varepsilon_j A_j
  T_j^4}_{P_\mathrm{radiated}}$$ 
where $J_{_\mathrm{L2}}$ is the solar constant, $\alpha_i$ the absorptance of each face, $A_{i/j}$ the area of 
each face, $T_j$ the temperature of each face and $\theta_i$ is the solar zenith angle of the faces. Only faces 
with a solar zenith angle within $\pm90^{\circ}$ contribute to the absorbed power. 

\noindent\textbf{Thermal design of the PM}\\
The PM consists mainly of the WFCs, the XFT, the IRT, a sunshield and radiators. The 30$^{\circ}$ cut 
(Figure\,\ref{fig:sc_geometry}, i.e. right side is smaller than the left side) is designed to shield the 
instruments from solar radiation. The XFT detector plane will reside in the SM, and so must be thermally 
controlled there. The IRT has to be kept on the shaded side in order to avoid heat as much as possible, because 
of its sensor temperature requirement of $\sim$\,35\,K. Thermal insulation has to be optimal, with a maximum 
heat transfer, Q, of 1 W. The WFCs, which consume a lot of power, are kept at room temperature. 
Table~\ref{tab:thermal_pm} provides an overview of the relevant instrument data.

\begin{table}[ht]
\scriptsize{
\centering
\begin{tabular}{lcccc} \toprule
Instrument 	&Power dissipation & Temperature	& Heat Transfer	& Area required     	\\ 
		&(W)		&(K)		&(W)		&(m$^2$)	\\ \midrule
WFC 		&200		&293		&10 		&0.03		\\
IRT 		&20		&35		&1 		&13.06		\\ 
XFT 		&15		&253		&2 		&0.01		\\ \bottomrule
\end{tabular} 
\caption{Overview of instrument details relevant for thermal engineering. The area required is that 
necessary to cool the instrument to the desired temperature.} 
\label{tab:thermal_pm}
}
\end{table}

\begin{figure}
        \centering
        \subfigure[PM, surface is at 253 K \label{fig:pm_thermal_final}]{%
        \includegraphics[width=6cm]{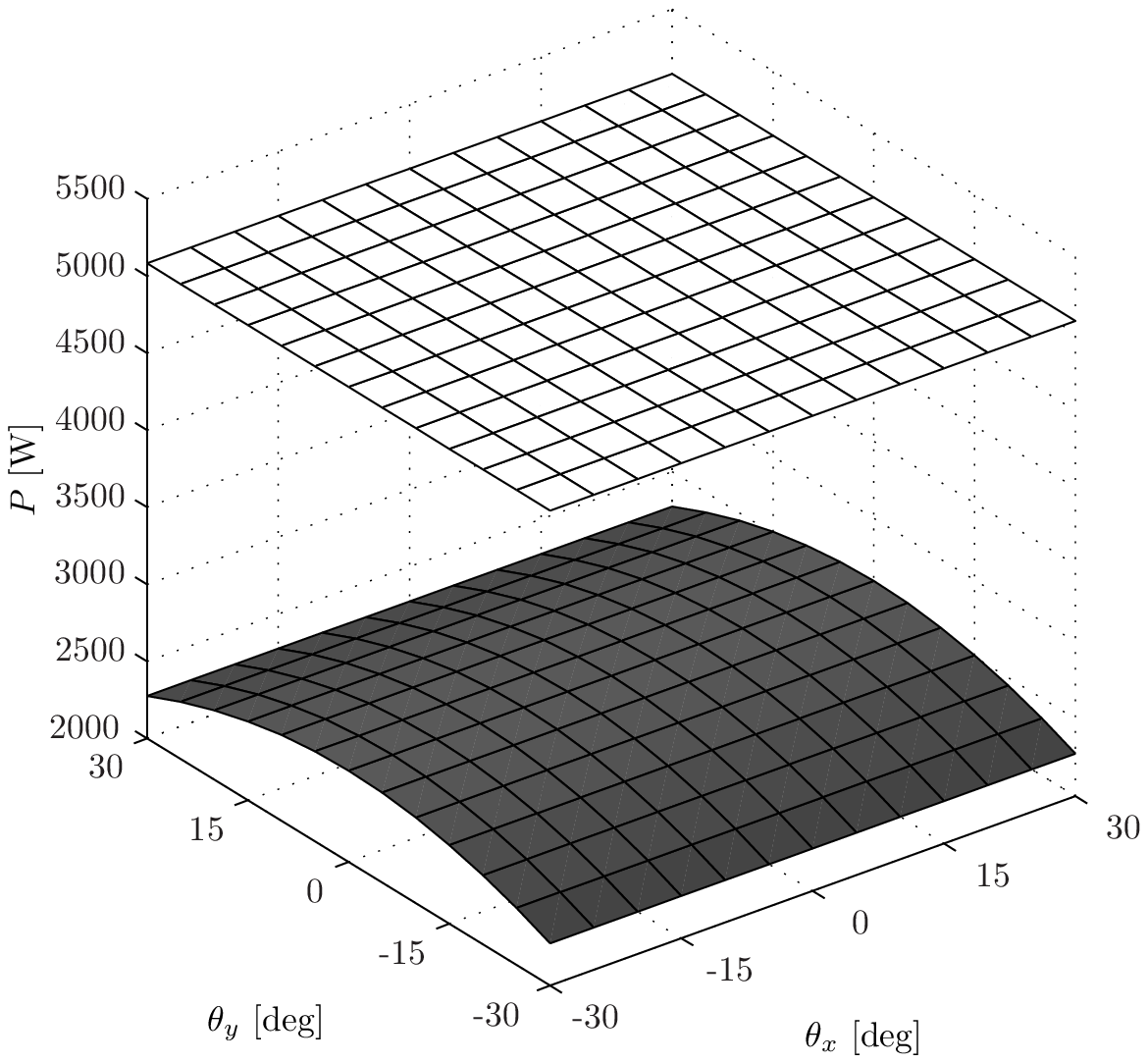}
        }%
        \hspace{0.5cm}
        \subfigure[SM, surface is at 293K \label{fig:sm_thermal_final}]{%
        \includegraphics[width=6cm]{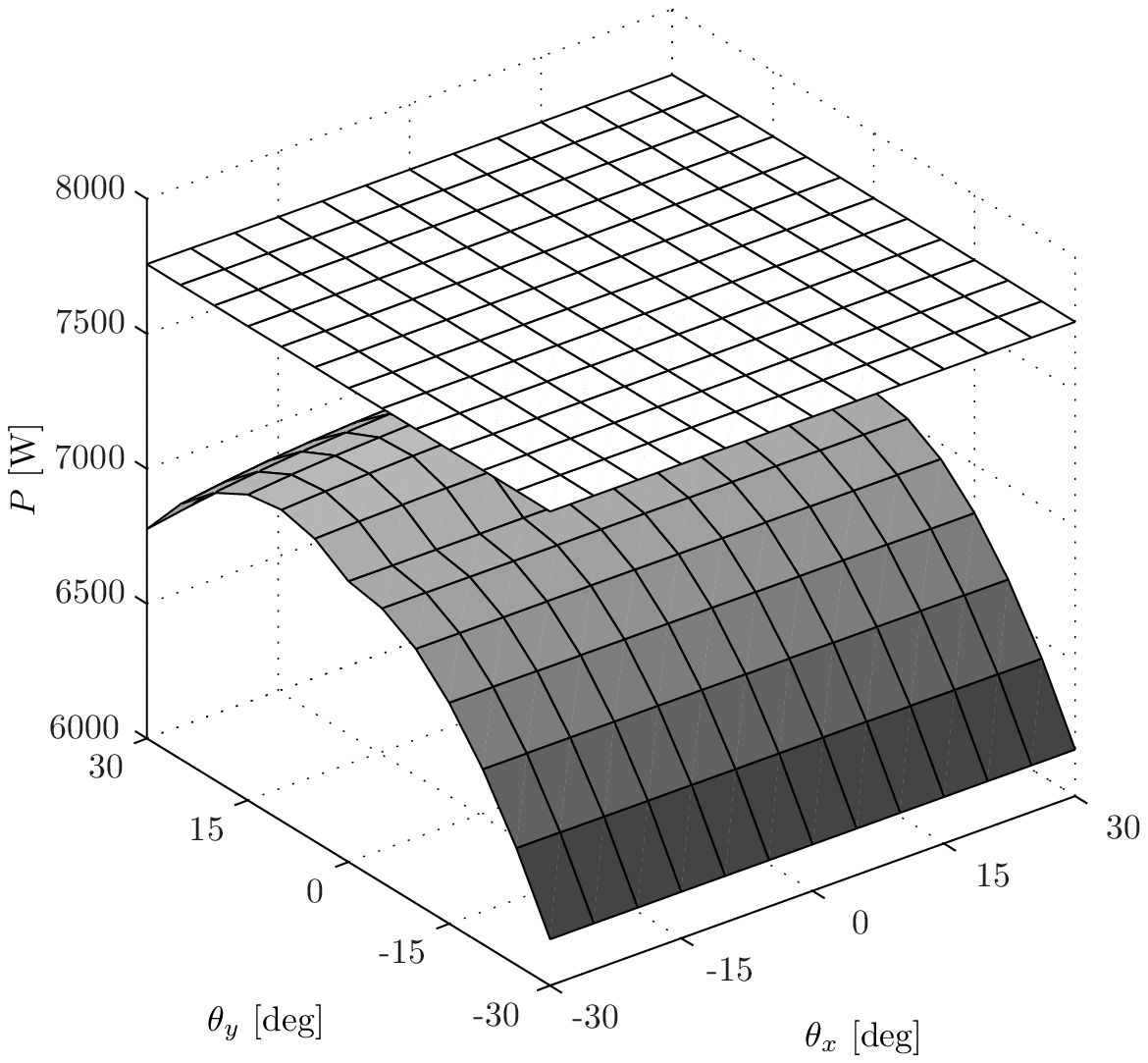}
        }
        \caption{Thermal PM and SM performance. The plane surface represents the emitted power,
the curved surface is
the absorbed power plus the electrically dissipated power.}
        \label{fig:thermal_final}
\end{figure}

Parts of the payload structure that are facing the Sun are coated with white paint (e.g. Alion~products YB-71 and YB-71P 
with $\alpha$\,$\sim$\,0.12 and $\varepsilon$\,$\sim$\,0.90), the back side is passively cooled by radiating areas 
coated with black paint (e.g. Alion~products MH216NLO and MH21-IP with $\alpha$\,$\sim$\,0.95 and 
$\varepsilon$\,$\sim$\,0.90). 
Therefore, the required area follows from 
$$A_j=\frac{Q_j}{\sigma \varepsilon T_j^4}$$
where $Q_j$ is the heat transfer for the $j$th area and $T_j$ the corresponding temperature. 
Simulations showed that the required area for the radiators (Table\,\ref{tab:thermal_pm}) does not exceed the 
available area. This is of particular interest since 35~K for the IRT is difficult to achieve by passive 
cooling. By using the suggested material the thermal budget yields more radiated power than absorbed power as 
can be seen in Figure~\ref{fig:pm_thermal_final} because the white paint covers a rather big area and has a 
high emittance. This fact is important because heating is always easier than cooling. By using different 
materials (polished aluminium or black paint) or heaters, a more balanced thermal budget is possible. In 
our analysis, we neglected detailed structures, such as the high gain communication antenna dish on the front, 
which would contribute to heating rather than to cooling. \\

\noindent\textbf{Thermal design of the SM}\\
The front face of the SM is completely covered by solar panels, which will heat up the module. The thermal
design of the SM focusses mainly on keeping the module at room temperature to guarantee a stable operating 
temperature for its devices, particularly the batteries and the propellant. This is done by radiators on the 
rear face. Given that there is also an instrument in the SM, the XFT, an additional radiator panel has to be 
provided to keep it at the specified temperature (Table~\ref{tab:thermal_pm}). Simulations confirmed that, 
within rotations of $\pm 30^{\circ}$ (about the x and/or y axis), the radiated power always exceeds the sum of 
the absorbed and dissipated power (Figure~\ref{fig:sm_thermal_final}).

In summary, it is shown that the proposed design is feasible in providing the required thermal specifications 
of all instruments and devices. The layout allows operation of the IRT in the temperature range of 35-50 K. 
Calculations reveal the engineering challenge to be insulation of the IRT such that the heat transfer does not 
exceed 1\,W, in which case it will not be possible to allocate sufficient area for the radiators. 

\subsection{Telemetry}
\label{sect:telemetry}

The aim of this section is to estimate the amount of data produced by the various on-board instruments, and to examine 
the best way to communicate with the ground stations. The largest amount of data will undoubtedly 
come from the WFCs. This is mainly due to the diffuse x-ray background, the intensity of which is estimated 
as follows, \cite{Gruber1992}, \cite{Revnivtsev2003}. The number of photons in a given energy range 
(\,cm$^{-2}$\,steradian$^{-1}$\,s$^{-1}$) is given by: 
\begin{equation}
\frac{\mathrm{d} N(E)}{\mathrm{d} E}=9.8 \frac{\mathrm{phot}}{\mathrm{s\,cm^2\,keV\,sr}} E^{-1.42}\ 
\hspace{1cm}\mathrm{for} \,\,1\,\mathrm{keV}<E<15\,\mathrm{keV} \\ \nonumber
\end{equation}
\begin{equation}
\frac{\mathrm{d} N(E)}{\mathrm{d} E}=167 \frac{\mathrm{phot}}{\mathrm{s\,cm^2\,keV\,sr}} E^{-2.38}\ 
\hspace{2.2cm}\mathrm{for} \,\,E>15\,\mathrm{keV} \nonumber
\end{equation}
Integrating over the appropriate energy range (0.1-100~keV), and multiplying by the total detector area and 
FOV of the four WFCs, gives a total number of 87\,000\,photons\,s$^{-1}$ (assuming coded mask efficiency 
$\sim$\,50\,\%). The relevant data that needs to be stored for every photon consists of six characterising 
numbers: the (x,y) position on both detectors, the x-ray energy, and the time. Since each of these will be 
stored on 10\,bits, it amounts to 60\,bits per photon. So we expect the data rate of the 4 WFCs to be 
$\sim$5.3\,Mb\,s$^{-1}$, while the XFT and IRT require comparatively negligible rates (i.e. 
$\sim$50\,kb\,s$^{-1}$). The total comes to $\sim$5.5\,Mb\,s$^{-1}$. 

The large data rate leaves two alternative. The preferred scenario involves using a high-gain, steerable 
antenna (HGA) in searching for GRBs. In addition to the HGA continuously transmitting WFCs data to Earth, a 
medium-gain antenna (MGA) ensures minimal data transmission for housekeeping and emergency situtations. Also, 
the spacecraft must be capable of storing data onboard for a limited time. Estimations show the need for a 
1.2\,m diameter HGA and a ground antenna diameter of 30\,m. However, various restrictions have to be considered. 
Currently, available ground stations cannot ensure complete ground coverage, and a steerable HGA can constrain 
the observable region. However, the data rate is the main concern, even given data compression. For 
other space missions, data transmission rates of almost 3\,Mb\,s$^{-1}$ are under development and will be 
achievable from L2 when satellite missions like GAIA \cite{Lammers2005} are launched. The alternative 
scenario uses a MGA, which inevitably means onboard raw-data processing and storage. The 
reduced data are transmitted to ground during a limited time interval (typically a few hours). The first 
scenario is much more attractive, mainly because of the higher flexibility and  real-time ground-based data 
processing, but its feasibility depends on future developments mainly in antenna technology. 

\section{Mass Estimate}
\label{mass_estimate}

One of the key points to note about this mission design is the light-weight technology and feasibility of 
passive cooling (which does not increase mass through fluid considerations). This allows us to choose a 
smaller launcher (Section\,\ref{orbit}), which ultimately leads to a significant reduction in overall mission costs. The 
individual masses for the subsystems are roughly broken down as follows: WFC$\sim$160\,kg ; XRT $\sim$90\,kg; IRT 
$\sim$150\,kg; attitude control system $\sim$114\,kg; power system \& telemetry $\sim$130\,kg; structure 
$\sim$550\,kg. This gives a total mass of the spacecraft within the mass limitations for a Soyuz launcher, 
i.e. 1\,500\,kg.

\section{Observing Strategy} 
\label{observing_strategy}

Observing the sky for GRBs is the primary objective of the {\bf\em X-red} mission. The mission design accounts for the 
fact that we cannot observe in either the galactic or ecliptic planes, so as to avoid high x-ray background levels and 
possible x-ray flashes unrelated to GRB events. 

While the WFCs are monitoring the sky for GRBs, the WFCs data (with positional accuracy of 1 arcmin) are analyzed by the 
satellite's software and are continuously transmitted to Earth in real time. The spacecraft then acts automatically on the 
coordinate information, to take further data with the other instruments. However, the satellite's GRB detection software 
will have a high detecting threshold, to prevent triggering by false events. Ground stations have the option to analyze 
the WFCs' data, and may choose to override the satellite controls in order to study GRBs that failed to reach the 
detection threshold. In this case, the total time loss would be 20\,s. Whether by automatic instruction or manual 
override, the satellite will slew to point the XFT and IRT on the GRB position in less than 60\,s. The XFT and IRT 
instruments will then observe the GRB and its afterglow. The XFT can improve the positional accuracy to within 1 
arcsecond, and immediately transmit this information to Earth where ground-based telescopes may choose to begin follow-up 
observations. Note, however, that thermal and power issues place restrictions on spacecraft orientation and so, following 
observation of several GRBs, the spacecraft must be reorientated to a position that ensures optimal thermal and power 
conditions. 

\section{Conclusions}

Extremely high redshifted GRBs provide unique probes of the first generation of stars and the early Universe. 
We present {\bf\em X-red}, a complete mission design from the science case to the payload and overall mission 
architecture, the objective of which is to observe these high redshift GRBs for the first time. 

We have demonstrated the feasibility of detecting high redshift GRBs in terms of both statistical number of 
detections and also adequate flux levels at x-ray and infrared wavelengths. The results dictated the demands 
on instrumentation. We then describe the payload which comprises three instruments, namely wide field x-ray 
cameras to detect high redshift gamma-rays, an x-ray focussing telescope to determine accurate coordinates and 
extract spectra, and an infrared spectrograph to observe the high redshift optical afterglow. We successfully 
demonstrate that our scientific objectives are achieveable when harnessing new technologies (i.e. DEPFET and pore optics), 
and we successfully meet instrument accomodation and environmental demands through our mission architecture. One of the 
key points to note about our overall mission design is the light-weight technology and feasibility of passive cooling. 
This significant reduction in mass allows us to choose a smaller launcher, which ultimately leads to a significant 
reduction in overall mission costs {\em without} compromising scientific objectives. 

In the planned 3 years mission, we expect to detect and identify 15 GRBs of z$>$10. These will constitute the 
first observations of high redshift GRBs, and are expected to provide valuable constraints on properties of 
the first generation of stars and on the history of the early Universe. 

\subsection{Acknowledgments} 
We would like to thank all the tutors of the 2004 ESA summer school 
for their help and suggestions, our dear friend Prof. Dr. 
Johannes Ortner 
for a perfectly organized summer school, all professors for their 
inspiring lectures, the summer school's staff,  ESA and its 
single national societies (DSRI, DLR, ASA, etc.) for the financial
support of the summer school and participants. 
Without this help, the collaboration for this paper would not have been 
possible.

{\small Mirko Krumpe is supported by the Deutsches Zentrum f\"ur Luft-
und Raumfahrt (DLR) GmbH under contract No. FKZ 50 OR 0404.}

\newpage
\begin{figure}[h]
 \begin{center}
 \epsfig{file=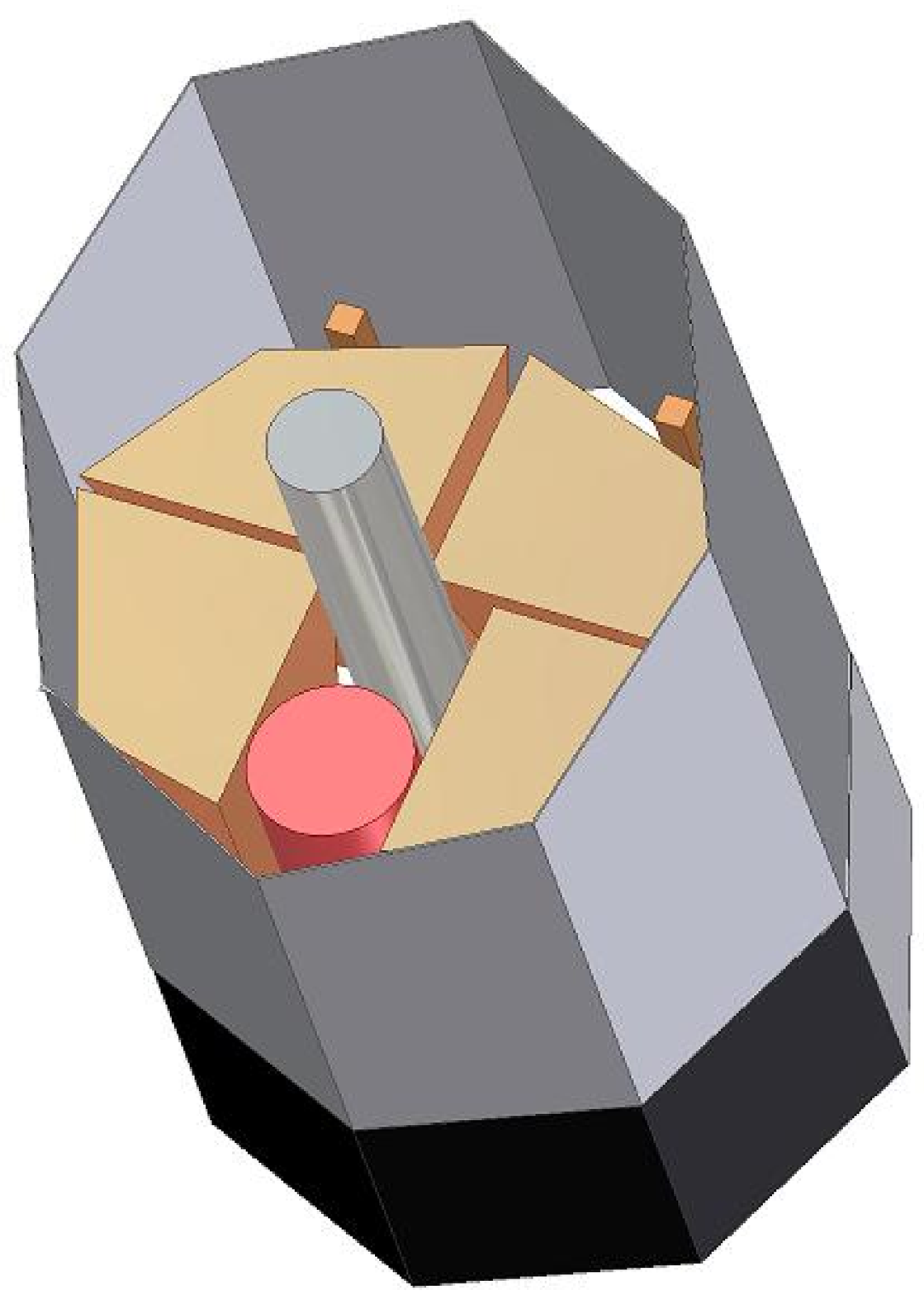,width=10cm,angle=0}
 \caption{X-RED}
 \label{f:absorption}
 \end{center}
\end{figure} 


\bibliographystyle{spiebib}   

\newpage

\end{document}